\documentclass[12pt]{iopart}

\usepackage{graphicx}	
\usepackage{subfigure}
\usepackage{amssymb}	
\usepackage{mathptm}	
\usepackage{color}	

\newcommand{\be}{\begin{equation}}
\newcommand{\ee}{\end{equation}}
\newcommand{\bea}{\begin{eqnarray}}
\newcommand{\eea}{\end{eqnarray}}

\begin{document}


\title[Universality in statistical measures of classical trajectories]{Universality in statistical measures of trajectories in classical billiards: Integrable rectangular versus chaotic Sinai and Bunimovich billiards}

\author{
	Jean-Fran\c{c}ois~Laprise$^{1,2}$,
	Ahmad~Hosseinizadeh$^{1}$,
	Helmut~Kr\"{o}ger$^{1}$ and
	Reza~Zomorrodi$^{1}$
}

\address{$^{1}$ D\'{e}partement de physique, Universit\'{e} Laval,
Qu\'{e}bec, QC, G1V~0A6, Canada}
\address{$^{2}$ Unit\'e de recherche en sant\'e des populations, CHA universitaire de Qu\'ebec, Qu\'ebec, QC, G1S~4L8, Canada}
\ead{\mailto{jlaprise@uresp.ulaval.ca}}


\begin{abstract}
For classical billiards we suggest that a matrix of action or length of trajectories in conjunction with statistical measures, level spacing distribution and spectral rigidity, can be used to distinguish chaotic from integrable systems. As examples of 2D chaotic billiards we considered the Bunimovich stadium billiard and the Sinai billiard. In the level spacing distribution and spectral rigidity we found GOE behaviour consistent with predictions from random matrix theory. We studied transport properties and computed a diffusion coefficient. For the Sinai billiard, we found normal diffusion, while the stadium billiard showed anomalous diffusion behaviour. As example of a 2D integrable billiard we considered the rectangular billiard. We found very rigid behaviour with strongly correlated spectra similar to a Dirac comb. These findings present numerical evidence for universality in level spacing fluctuations to hold in classically integrable systems and in classically fully chaotic systems.
%
\end{abstract}


\submitto{\JPA}

\maketitle



\section{Introduction}
\label{sec:Intro}
The idea to model apparently disordered spectra, like those of heavy nuclei,
using random matrices was suggested in the mid-50's by Wigner, and then 
formalized in the early 60's in the work of Dyson and Mehta~\cite{Mehta60,Dyson62,Mehta91}. 
They showed that random matrices of Gaussian orthogonal ensembles (GOE) 
generate a Wigner-type nearest-neighbour level spacing (NNS) 
distribution~\cite{Mehta91}. In a seminal paper, Bohigas, Giannoni and 
Schmit (BGS) formulated a conjecture~\cite{Bohigas84} stating that 
time-reversal invariant quantum systems with classically 
fully chaotic (ergodic) counterpart have universality properties given 
by random matrix theory (RMT).
Experiments in nuclear physics, for example, have shown that spectra 
originating from different heavy nuclei give the same Wignerian energy 
level spacing distribution~\cite{Bohigas83}. 
Universality properties in quantum chaos of bound systems, i.e. quantum systems
with a fully chaotic classical counterpart, have now been demonstrated in 
many experiments, computational models and in theoretical 
studies~\cite{Mehta91,Casati80,Blumel97,Stockmann99,Guhr98,Haake01,Robnik86}.
Theoretical support of the BGS conjecture came from the semiclassical theory of spectral rigidity 
by Berry~\cite{Berry85,Berry87}, who showed that universal behaviour in 
the energy level statistics is due to long classical orbits. 
Sieber and Richter~\cite{Sieber01} investigated the role of correlations 
between classical orbits. The semiclassical theory has been further 
developed by M\"uller et al.~\cite{Muller04a,Muller04b,Muller04c}.
In ref.~\cite{Muller04c} they presented the \textquoteleft core of a proof\textquoteright\ of 
the BGS conjecture, which, by proving arguments previously used by Berry~\cite{Berry85}, 
show that in the semi-classical limit the periodic classical orbits determine the 
universal fluctuations of quantum energy levels. Further refinements have been 
made by Keating and M\"uller~\cite{Keating07}.

The study of classical strongly chaotic systems (Anosov systems) has revealed 
that central limit theorems (CLT) 
hold~\cite{Bunimovich81,Burton87,Bunimovich91,Chernov94,Liverani95,Chernov95}. 
This has been proven for the 2D periodic Lorentz gas with finite horizon. 
The first step of a proof was given by Bunimovich and Sinai~\cite{Bunimovich81} and was completed by 
Bunimovich, Sinai and Chernov~\cite{Bunimovich91}.  
At macroscopic times, such deterministic chaotic system converges to Brownian motion, 
i.e. behaves like a random system~\cite{Bunimovich81,Bunimovich91,Chernov94,Chernov95}. 
This is also supported by the existence of an average diffusion 
coefficient~\cite{Bunimovich81,Bunimovich91,Gaspard98}, 
showing the diffusive character of such chaotic system.  

Do classical fully chaotic systems also exhibit universality properties?
This question was addressed by Argaman et al.~\cite{Argaman93}, who showed that there is universal behaviour 
in 2-points correlation functions of actions of periodic orbits of classically chaotic systems. As examples they 
considered the deformed cat-map and the baker-map. 
This has been elaborated further in a number of studies by 
Dittes et al.~\cite{Dittes94}, Aurich and Sieber~\cite{Aurich94}, Cohen et al.~\cite{Cohen98}, Tanner~\cite{Tanner99}, 
Sano~\cite{Sano00}, Primack and Smilansky~\cite{Primack00}, Sieber and Richter~\cite{Sieber01} 
and Smilansky and Verdene~\cite{Smilansky03}. 
Argaman et al. started from the assumption that spectral fluctuations of chaotic quantum systems 
follow the predictions of RMT and they derived a universal expression for classical correlation functions of 
periodic orbits via Gutzwiller's semi-classical trace formula.
They concluded \textquotedblleft The real challenge, though, is to find out whether these action correlations can be 
explained on a completely classical level\textquotedblright.

An answer was recently proposed by Laprise et al.~\cite{Laprise08} who found universal behaviour in classical 2D billiards by looking at fluctuations in spectra of classical action/length matrices from billiard trajectories. They showed that one could distinguish chaos from integrability in classical systems using RMT and an analogue of the BGS-conjecture. In particular, they considered the Lima\c{c}on/Robnik family of billiards, which interpolates between the chaotic cardiod billiard and the integrable circular billiard. For the cardioid billiard, the level spacing distribution $P(s)$ and spectral rigidity $\Delta_{3}(L)$ were found to be consistent with the GOE behaviour predicted by RMT. For the interpolating case close to the circle, the behaviour approached a Poissonian distribution. The circular billiard itself was found to be very rigid and strongly correlated and yielded $P(s) \propto \delta(s-1)$. The jump in behaviour at the transition to the circle is associated with the corresponding change in the symmetry group. 

This article extends the results of reference \cite{Laprise08} in the following directions: 
(i) We consider the 2D rectangular billiard as another example of an integrable billiard. Compared to the 
circular billiard, this billiard has lower symmetry (no group property). Nevertheless, it displays strong spectral 
correlation and rigidity like the circular billiard.
(ii) We present numerical studies for other chaotic 2D billiards: the Sinai-billiard and the Bunimovich stadium 
billiard.
(iii) In order to understand the observed universal behaviour in chaotic billiards, we present arguments
linking such behaviour to CLTs, diffusive and random walk behaviour. In particular, we present a mathematically rigorous result on the distribution of length 
of trajectories.

The answers we found can be summarized as follows: 
For the 2D Sinai billiard and the 2D Bunimovich stadium billiard 
the level spacing distribution $P(s)$ and 
spectral rigidity $\Delta_{3}(L)$ are consistent with predictions of 
RMT (GOE behaviour), i.e. show universal 
behaviour. This behaviour is statistically the same as the one observed 
in quantum chaos (obtained from energy level spacing distributions).
The implication of these findings is that RMT not only represents well 
the statistical fluctuation properties of the energy spectrum of chaotic 
quantum systems, but also those of the length spectrum 
of chaotic classical systems. Moreover, statistical fluctuations obtained 
from spectra of action/length matrices clearly distinguish chaotic from 
integrable systems.

\section{ Length and action of trajectories } 
\label{sec:approach}
In classical systems, chaos information is encoded in trajectories.
According to the Alekseev-Brudno theorem~\cite{Alekseev81} 
the temporal length $t$ is related via 
the Kolmogorov-Sinai $K$ entropy to the information $I(t)$ in 
the segment of trajectory, 
\be
\lim_{t \to \infty} I(t)/t = K ~ .
\ee
This motivates us to look at the length of trajectories and its fluctuation properties.
Let $L(x(t),\dot{x}(t),t)$ denote the Lagrangian of a system, 
let $x^{\mathrm{traj}}(t)$ denote a solution (trajectory) of the Euler-Lagrange equations, with boundary points $x_{\mathrm{in}}\equiv x^{\mathrm{traj}}(t_{\mathrm{in}})$ and $x_{\mathrm{fi}}\equiv x^{\mathrm{traj}}(t_{\mathrm{fi}})$. Let  
\be\fl
\Sigma = S[x^{\mathrm{traj}}] = \int_{t_{\mathrm{in}}}^{t_{\mathrm{fi}}} \rmd t 
L(x(t),\dot{x}(t),t)\mid_{x=x^{\mathrm{traj}}} ~ ,
\ee
denote the action over $x^{\mathrm{traj}}(t)$ and let 
\be
\Lambda[x^{\mathrm{traj}}] = \int_{s=x_{\mathrm{in}}}^{s=x_{\mathrm{fi}}} \rmd s\mid_{x=x^{\mathrm{traj}}} 
\ee
denote the length of the trajectory $x^{\mathrm{traj}}(t)$. 
We choose a finite set of discrete points 
$X=\{x_{1},x_{2},\dots,x_{N}\}$. For all pairs of boundary points 
$x_{i},x_{j}\in X$, , we compute a classical trajectory 
$x^{\mathrm{traj}}_{i,j}$, connecting those points.  
We suggest the construction of an action matrix and a length matrix,
\bea
\Sigma_{ij} = S[x^{\mathrm{traj}}_{i,\,j}] ~ ,
\nonumber\\
\Lambda_{ij} = \Lambda[x^{\mathrm{traj}}_{i,\,j}] ~ ,
\eea
where $i$ and $j$ are respectively the indices of the final and initial boundary points of the trajectory. Both matrices are viable for statistical analysis of 
classical chaos. In the case of billiard systems, we consider trajectories where 
the billiard particle moves with constant velocity $u$ and constant kinetic energy $E$. 
Then the action and the length matrix are essentially equivalent,
\be
\Sigma_{ij} = \frac{E}{u} \Lambda_{ij} ~ .
\ee
%


\subsection{Numerical calculation of length matrices}
\label{sec:matrix_construction}
When solving for two boundary points $x_i$ and $x_j$ in a billiard the solution is 
generally not unique; the number of possible trajectories varies with 
the number $N_{\mathrm{reb}}$ of rebounds. We have thus classified 
trajectories---and corresponding length matrices---according to $N_{\mathrm{reb}}$. 

For $N_{\mathrm{reb}}=15$ the number of trajectories is quite 
large ($N_{\mathrm{traj}} \sim 10^{5}$). In order to limit computational cost, 
we considered smaller subsets of length matrices, in the following way. 
For a given number of rebounds $N_{\mathrm{reb}}$ 
and a given pair of boundary points \{$x_{i}$, $x_{j}$\}, we determined 
the starting angles $\alpha_{\nu}$ (measured from the normal to the 
billiard wall at the boundary point) which corresponds to a trajectory 
$\nu$. That search was carried out in the range 
$-\pi/2 \le \alpha \le \pi/2$. 
The subsets have been constructed by 
introducing an upper bound $n_{\mathrm{traj}} (\ll N_{\mathrm{traj}})$ on the number of trajectories 
and retaining only the trajectories corresponding to the starting angles 
$\alpha_{1}, \dots, \alpha_{n_{\mathrm{traj}}}$.  
We repeated this for all combinations of boundary points.
We then checked that this procedure did not spoil the statistical properties we aimed to measure by considering different cutoff $n_{\mathrm{traj}}$ and different ordering scheme for the set of trajectories.

\section{Integrable billiard}
If one considers integrable quantum systems and analyzes them in 
terms of the NNS distribution of energy levels and spectral rigidity, then in 
most cases one finds a Poissonian distribution $P(s)$ which is 
reflected also in the behaviour of $\Delta_{3}(L)$.
However, there are examples of integrable quantum systems 
in 2D where the levels are correlated and the level spacing distribution is 
not given by a Poissonian. Such cases are called non-generic. 
Berry and Tabor~\cite{Berry77} noted as example 
uncoupled quantum oscillators in 2D. Casati et al.~\cite{Casati85} and later 
Seligman and Verbaarschot~\cite{Seligman86} showed that also the integrable 
2D quantum rectangular incommensurate billiard does not give an uncorrelated 
Poissonian level spacing distribution. In a recent study of 2D quantum 
harmonic oscillators Chakrabarti and Hu~\cite{Chakrabarti03} found in the 
case of uncoupled oscillators a level spacing distribution displaying a 
$\delta(s-1)$ peak plus some background. They measured the spectral correlation 
via the spectral rigidity $\Delta_{3}(L)$ and observed a flat curve for 
$L \ge 4$ saturating at $\Delta_{3}(L)|_{\mathrm{sat}}=0.17$. Similar results were 
found in the case with weak harmonic coupling, yielding $\Delta_{3}(L)|_{\mathrm{sat}}=0.12$ 
and correlation $C=0.997$. This behaviour is similar to the rigidity of a picket fence (Dirac comb) 
of equally spaced levels with $\Delta_{3}(L)=0.083$. They conclude that the 2D 
quantum oscillator system is highly correlated at short and long-range, is regular 
and very rigid.

Laprise et al.~\cite{Laprise08} considered the classical integrable circular billiard 
and constructed a length matrix from classical trajectories between boundary 
points located evenly on the billiard wall. They found 
saturation in the spectral rigidity for large $L$ at $\Delta_{3}(L) =0.11$ which 
behaves like the rigidity of a Dirac comb ($\Delta_{3}(L) =0.083$). 
In conjunction with a correlation coefficient of $C=0.999$ and a level spacing distribution
$P(s) \propto \delta(s-1)$ (Dirac comb) this indicates high correlation at short and 
long range and very rigid behaviour.

\subsection{Rectangular Billiard}
\begin{figure}[ht]
\centering
\includegraphics[width=6cm]{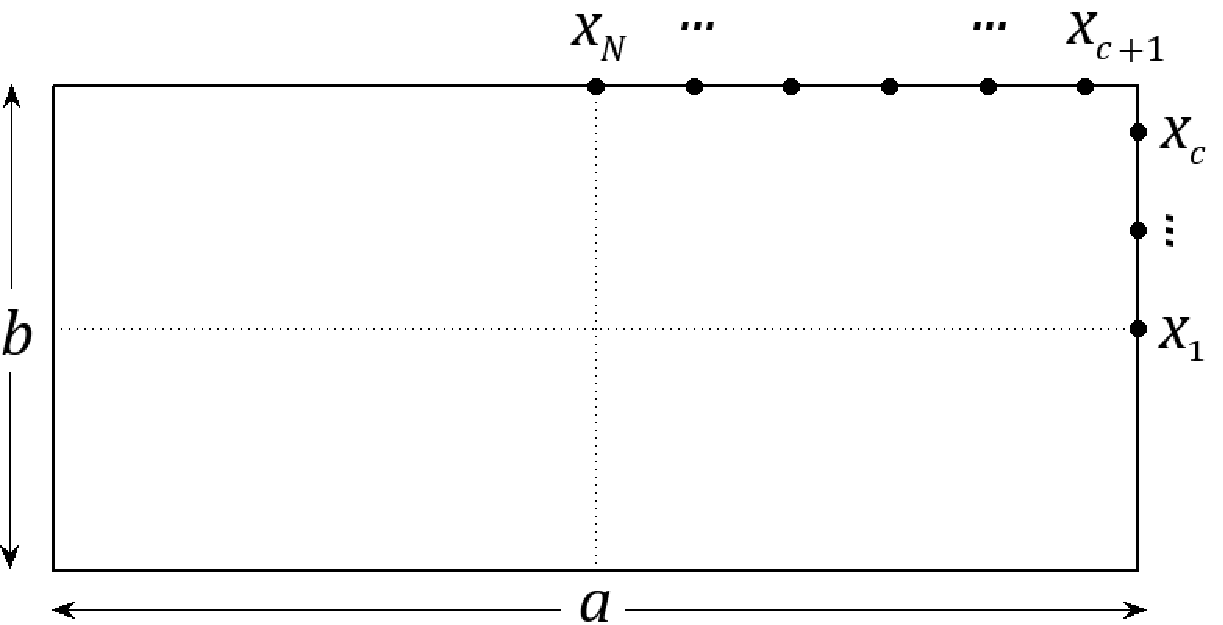}
\caption{2D Rectangular billiard. Horizontal length $a$, vertical length $b$. Trajectories go from boundary points $x_{i}$ to $x_{j}$ located in the upper right quarter of the  billiard wall.}
\label{fig:RectangleGeometry}
\end{figure} 
%
%
%
%
\begin{figure}[ht]
\centering
\subfigure[]{\label{fig:RectangleStatistics_top}
\includegraphics[width=6cm]{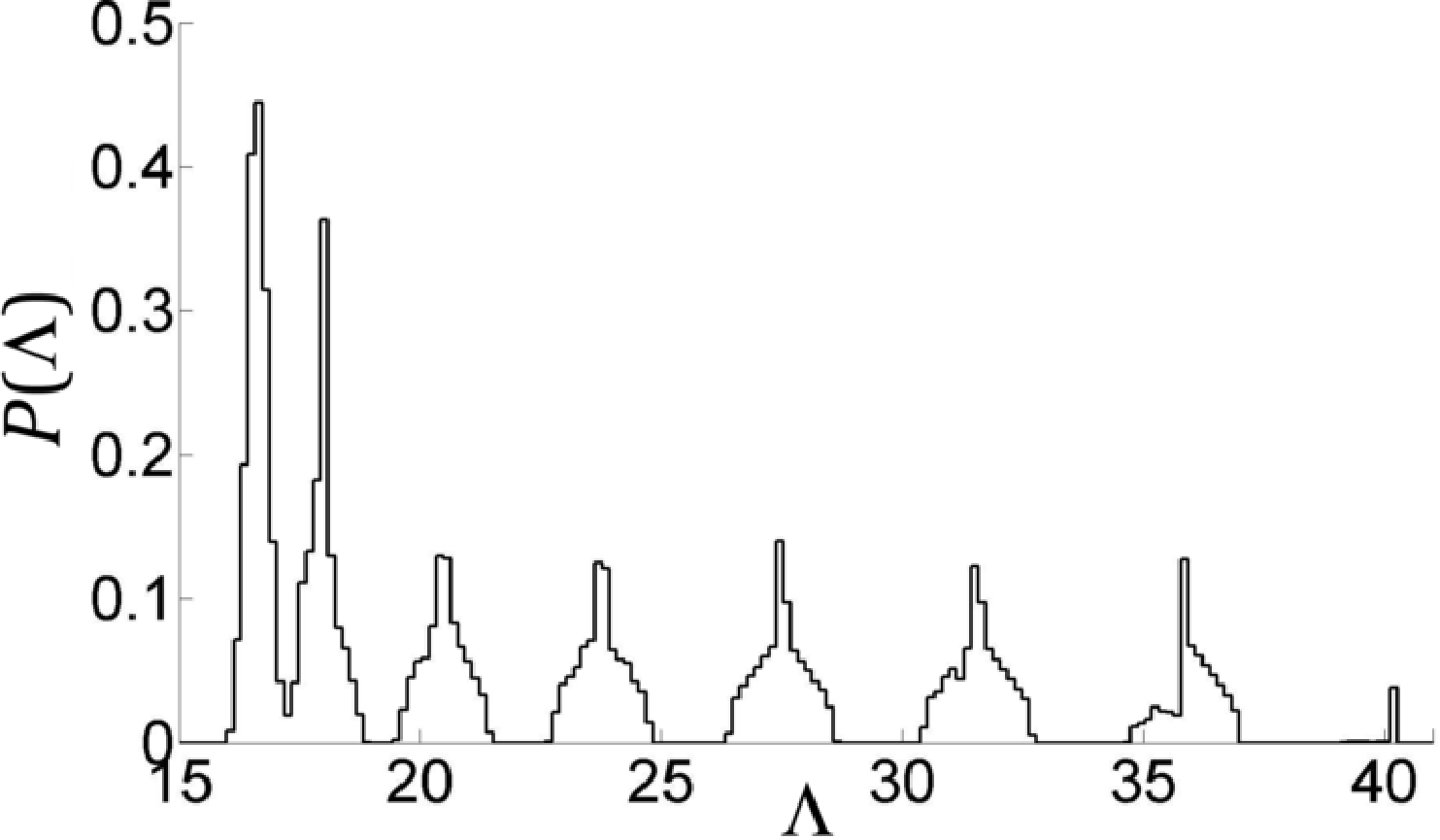}}\\
\subfigure[]{\label{fig:RectangleStatistics_middle}
\includegraphics[width=6cm]{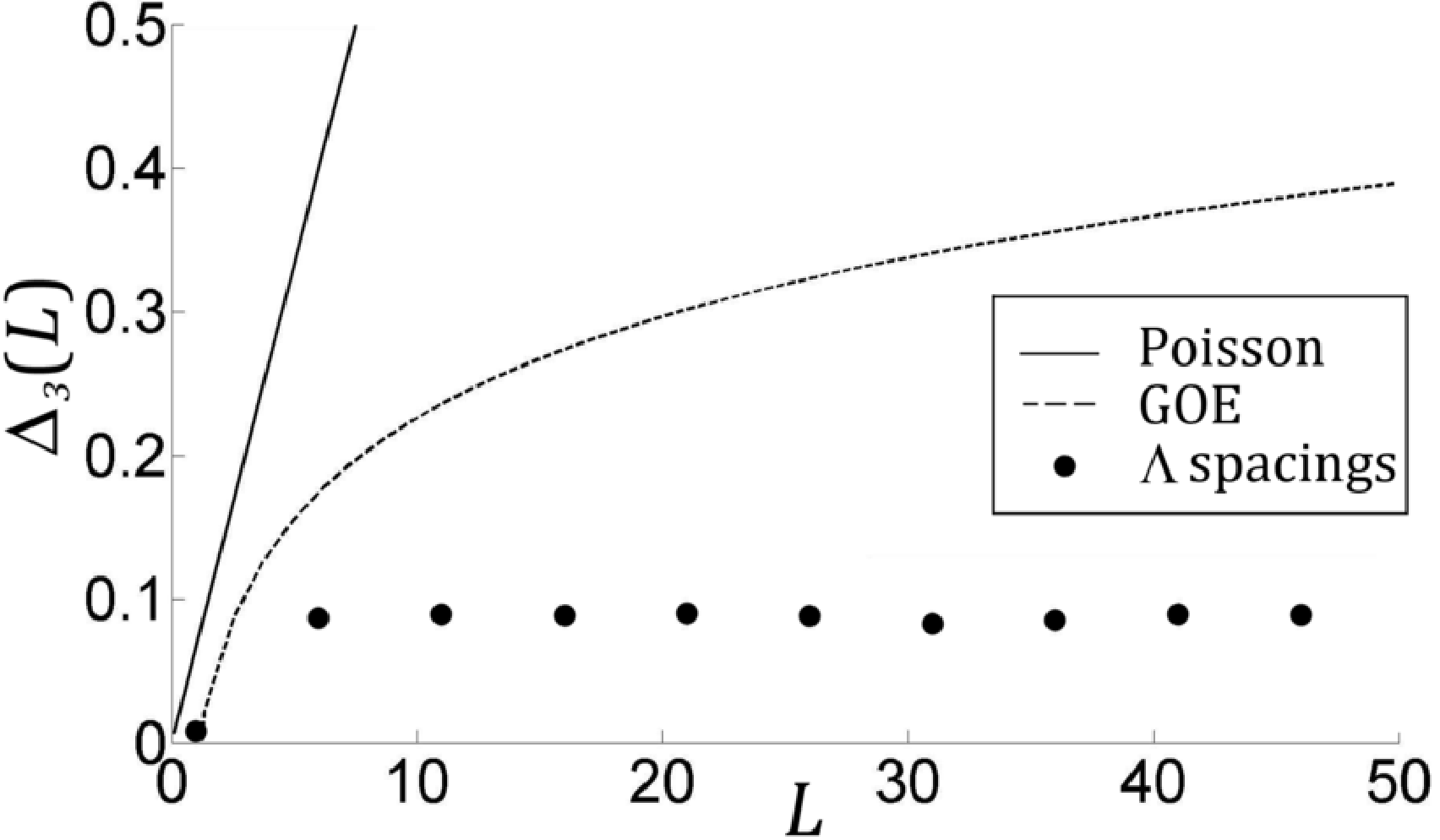}}\\
\subfigure[]{\label{fig:RectangleStatistics_bottom}
\includegraphics[width=6cm]{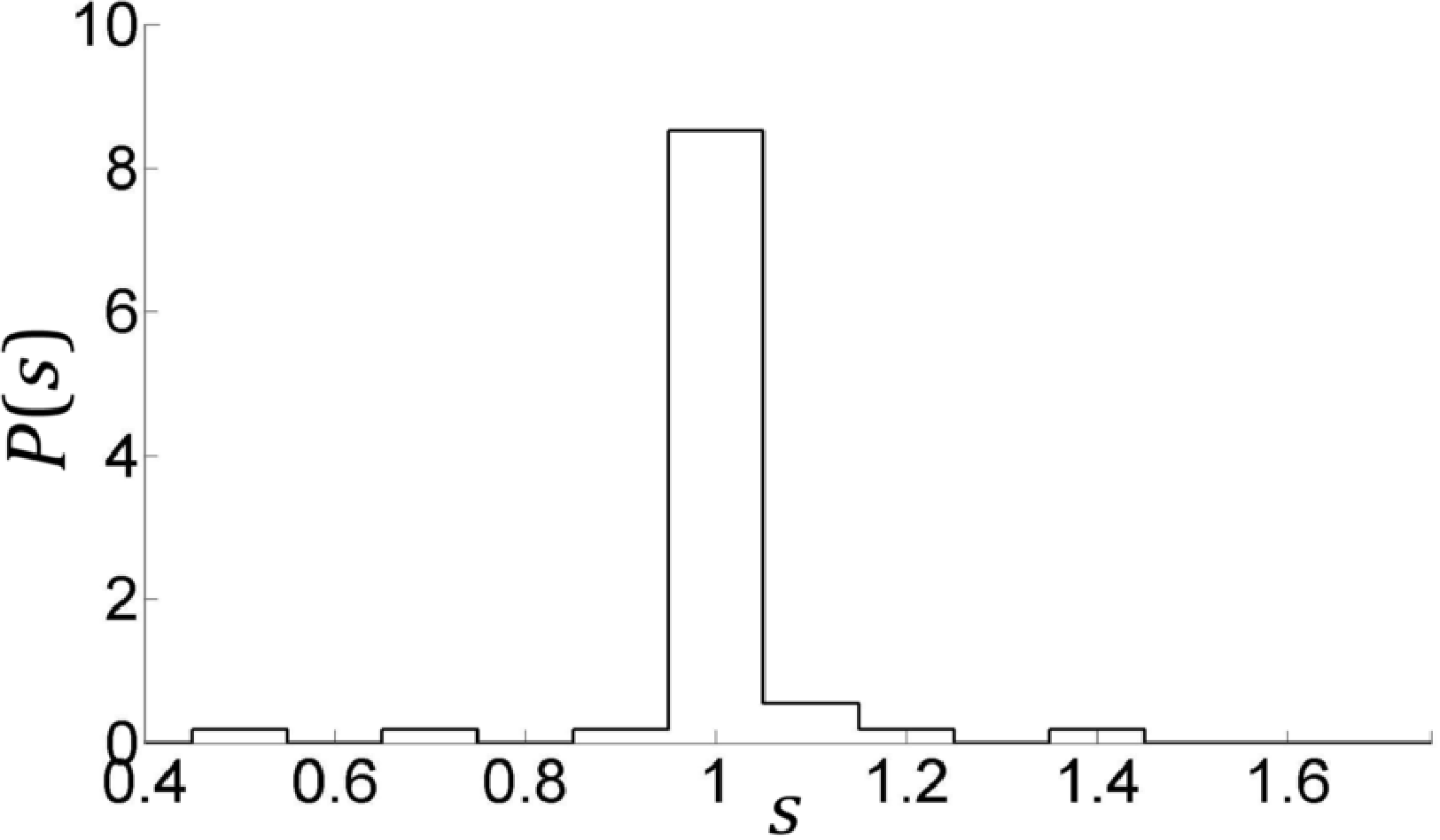}}
\caption{Rectangular billiard. 
(a) Distribution of length matrix elements $P(\Lambda)$. 
(b) Spectral rigidity $\Delta_{3}(L)$ of length spectrum.
(c) Level spacing distribution $P(s)$ of length spectrum.}
\label{fig:RectangleStatistics}
\end{figure}
As example of an integrable classical billiard we consider the 2D 
rectangular billiard, shown in figure \ref{fig:RectangleGeometry}. 
The shape is determined by the parameters $a$, $b$, which were chosen 
to be $a=\sqrt{5}$ and $b=1$. The boundary points 
$\{x_{1},\dots,x_{N}\}$, located in the upper right quarter of the 
billiard wall, are distributed regularly, with perimeter spacing given by
\be\fl
\label{eq:RectBillPerimSpac}
d(x_{n+1},x_{n}) = \frac{a+b}{2(N-1)} ~ , ~ n=1,\dots,N-1 ~ .
\ee
For a given pair of boundary points, 
we found that the behaviour of the number of trajectories versus the number of 
rebounds is linear (not shown). 
The error behaviour of trajectories as a 
function of the number of rebounds has been obtained by taking into account 
$N_{\mathrm{traj}}=20$ trajectories corresponding to different initial conditions. 
The error was stable in the regime $0 < N_{\mathrm{reb}} <100$ and 
did not increase beyond $10^{-16}$. This is in contrast to an exponential 
increase found in chaotic billiards. 

The symmetry of the rectangular billiard is mirror symmetry under reflection about the 
$x$- and $y$-axes (with origin at centre of rectangle). 
The symmetry shows up in the shape of trajectories. For example, a 
trajectory (1) going from starting point $x_{N}$ to endpoint $x_{N-1}$ bouncing 
once from the lower boundary wall has the same shape as the trajectory (2) going 
from starting point $x_{N-1}$ to endpoint $x_{N-2}$ bouncing once from the lower 
boundary wall. Trajectory (2) is obtained by a translation of trajectory (1) in 
$x$-direction by the amount of perimeter spacing, equation~(\ref{eq:RectBillPerimSpac}). 
However, such discrete translations do not 
form a group (they would form a group if one would consider the rectangular billiard 
with periodic boundary conditions at all billiard boundary walls, which would map the 
billiard on a torus, and the symmetry group would then be a rotation group).
The length matrix is not exactly a circular matrix either, however, it shares 
with circular matrices the property of having a number 
of different matrix elements equal to or less than the rank of the matrix. 
This property implies strong correlations between 
length matrix elements, which translates to strong correlations of eigenvalues of 
the length matrix. To sum up, although there is no 
exact symmetry group, there are residues of a \textquotedblleft broken symmetry group 
of discrete translations,\textquotedblright which imply strong correlations among length matrix 
elements and among eigenvalues of the length matrix.
We expect that this will manifest itself in the statistical behaviour
in the level spacing distribution $P(s)$ and the spectral rigidity $\Delta_{3}(L)$.
The results corresponding to the parameters $N=17$ and $N_{\mathrm{reb}}=55$ 
are shown in figure \ref{fig:RectangleStatistics}. 
figure \ref{fig:RectangleStatistics_top} represents the distribution 
$P(\Lambda)$ of length matrix elements. Its shape looks quite different from the 
near-Gaussian shape for chaotic billiards (see below).
figure \ref{fig:RectangleStatistics_middle} shows the spectral rigidity 
$\Delta_{3}(L)$ of the spectrum of the length matrix. It 
rapidly saturates to a value $\Delta_{3}(L)|_{\mathrm{sat}}=0.088$, which is close to the value 
$\Delta_{3}(L)|_{\mathrm{Dirac}}=1/12 \approx 0.083$ of an ideal Dirac comb. We have also 
computed the correlation coefficient to obtain $C=0.996$. This is consistent also 
with the NNS level distribution $P(s)$ of the length matrix shown in 
figure \ref{fig:RectangleStatistics_bottom} where one observes 
$P(s) \propto \delta(s-1)$ indicating the behaviour of a Dirac comb. 
The rectangular billiard turns out to be highly 
correlated at short and long range, to be very rigid and to be regular.

This is possibly evidence for universal 
behaviour in the integrable case. Comparing the behaviour of the rectangular billiard with the 
circular billiard~\cite{Laprise08}, we observe that they differ in their symmetry properties. In the 
circular billiard hopping from one boundary point to its neighbour stands for a group operation. 
The corresponding operation in the rectangular billiard has no group property. 
However, the resulting strong correlation and spectral rigidity are found to be very similar for both billiards.

\section{Chaotic billiards}
\subsection{Sinai Billiard}
\begin{figure}[ht]
\centering
\subfigure[]{
	\label{fig:SinaiGeometry}
	\includegraphics[width=3.5cm]{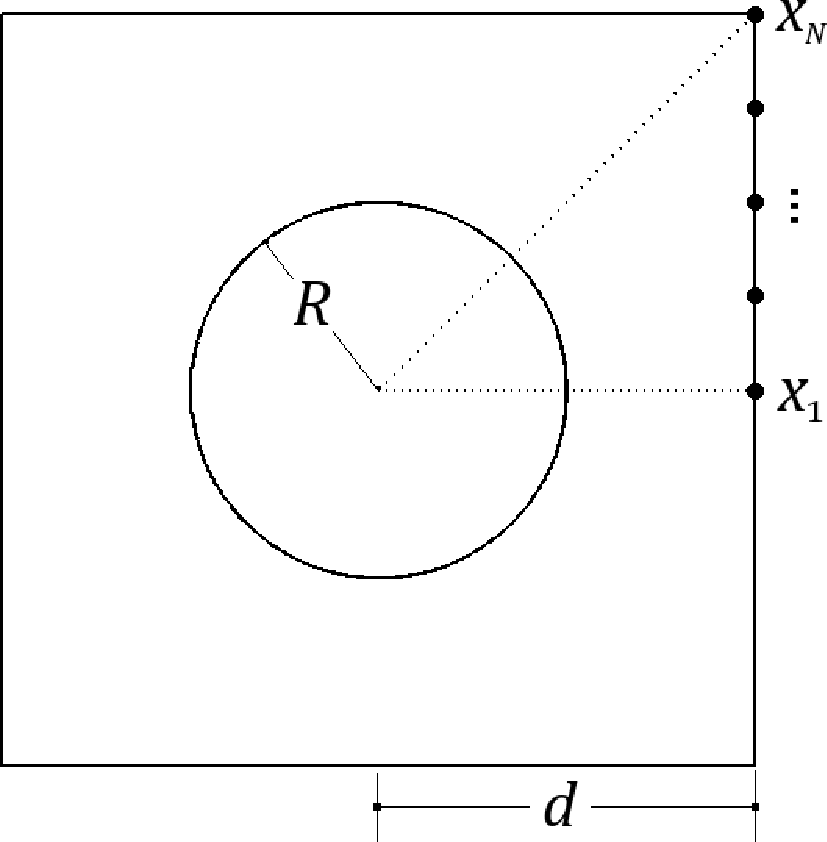} 
}\\
\subfigure[]{
	\label{fig:SinaiNumSol+Error}
	\includegraphics[width=6cm]{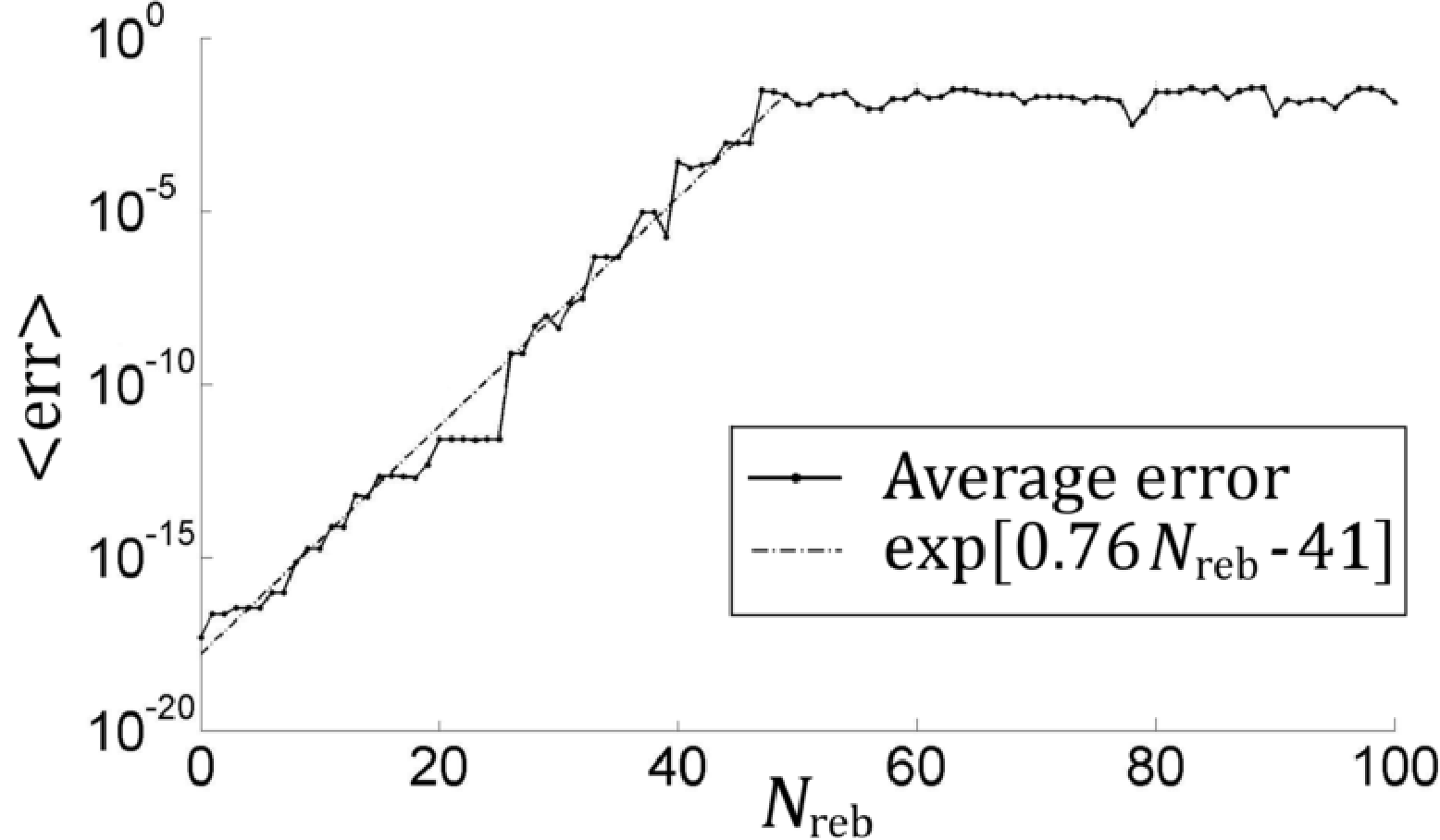}
}
\caption{
(a) Geometry of 2D Sinai billiard. Parameters: radius of sphere $R=1$, half-length of box $d=5$. Trajectories go from boundary points $x_{i}$ to $x_{j}$ located in a segment 
of one eighth of the wall of the billiard. (b) Relative error as function of number of rebounds for 2D Sinai billiard.
}
\end{figure} 
%
%
%
%
%
%
%
\begin{figure}[ht]
\centering
\subfigure[]{\label{fig:SinaiLength+LSD+Delta3_top}
\includegraphics[width=6cm]{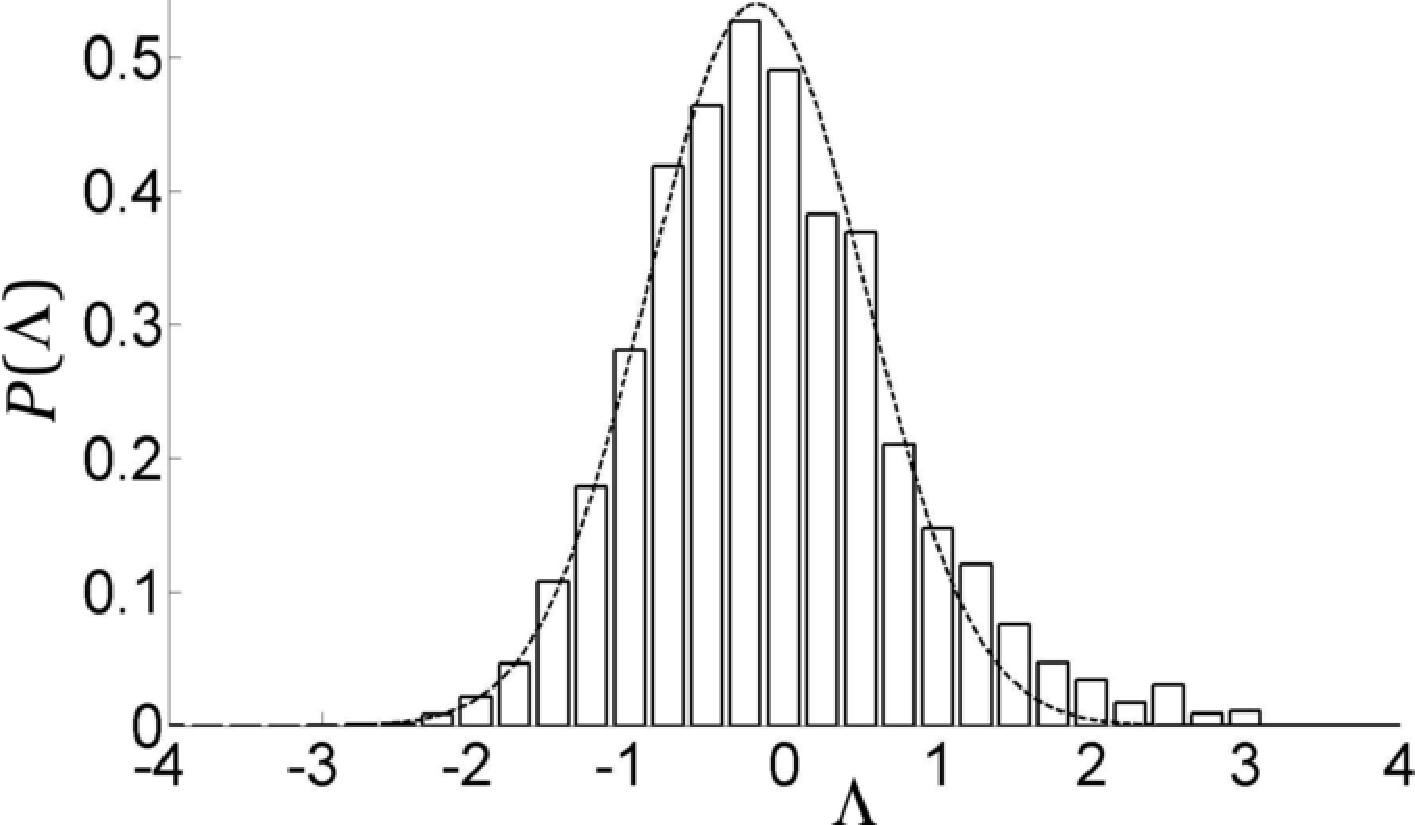}}\\
\subfigure[]{\label{fig:SinaiLength+LSD+Delta3_middle}
\includegraphics[width=6cm]{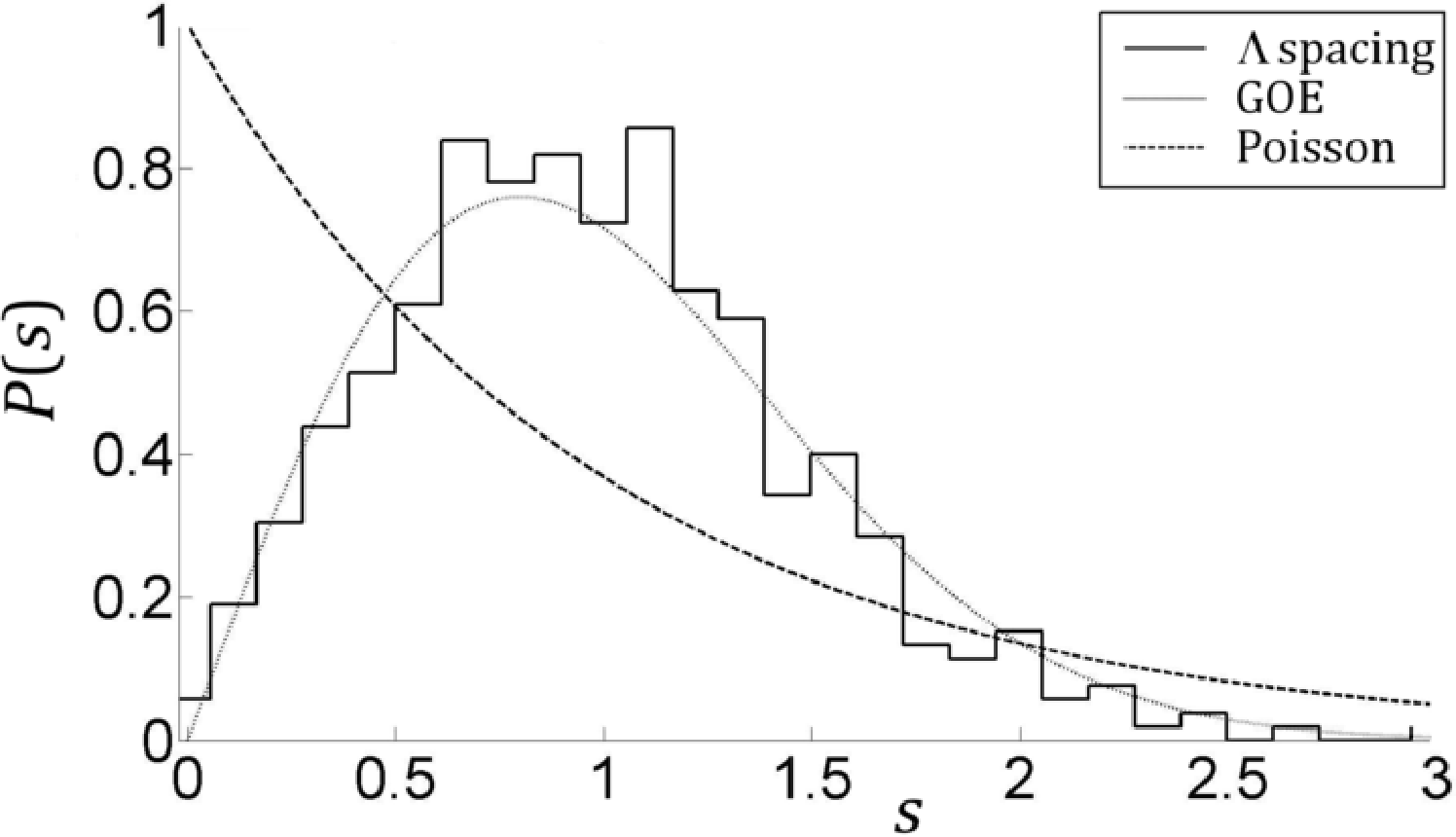}}\\
\subfigure[]{\label{fig:SinaiLength+LSD+Delta3_bottom}
\includegraphics[width=6cm]{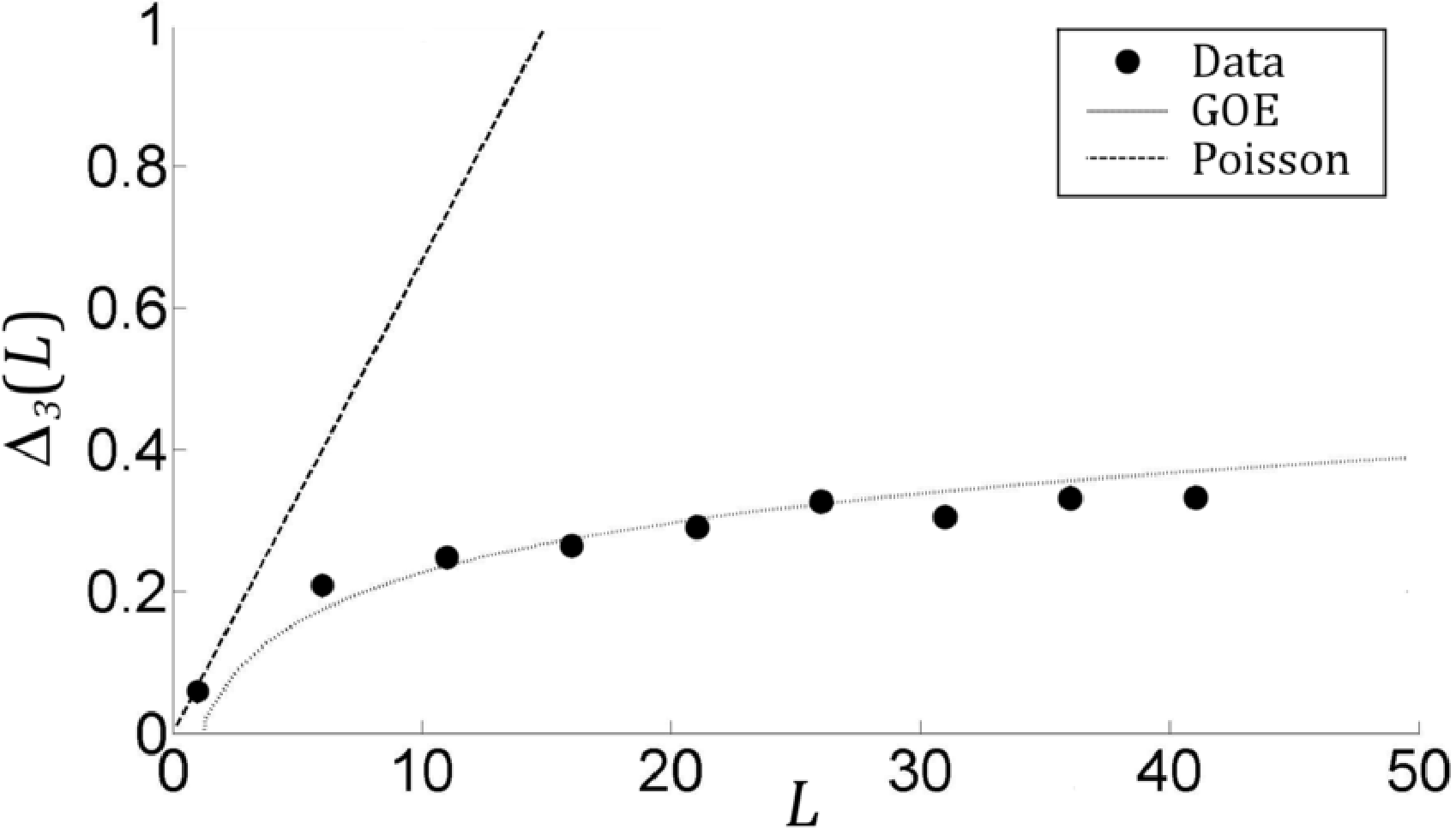}}
\caption{2D Sinai billiard. $N_{reb}=15$. 
(a) Distribution of length matrix elements (after subtracting mean $<\Lambda>$).
(b) Level spacing distribution from length matrix $\Lambda$.
(c) Spectral rigidity.}
\label{fig:SinaiLength+LSD+Delta3}
\end{figure}
For general closed 2D billiards, the mean free path length $\Lambda_{\mathrm{free}}$ in between 
two collisions is given by the billiard geometry via~\cite{Chernov97}
\be
\label{eq:FreePathLength}
\Lambda_{\mathrm{free}} = \frac{\pi |Q|}{|\partial Q|} ~ ,
\ee
where $|Q|$ stands for the billiard area and $|\partial Q|$ for the length of the wall.
Let us now consider a particle of mass $m$ moving in the 2D Sinai 
billiard (see figure \ref{fig:SinaiGeometry}).
The classical Sinai billiard system is known to be fully chaotic~\cite{Sinai63}. 
The billiard is symmetric under mirror operation about $x$- and $y$-axes, 
and about the diagonals  of angles $\pi/4$ and $3\pi/4$, respectively, 
passing through the centre. In order not to mix different symmetry 
classes, we consider the billiard with boundary points $x_{1},\dots,x_{N}$ 
located in one eighth of the exterior billiard wall 
(see figure \ref{fig:SinaiGeometry}). It is important to make sure 
that the chaotic behaviour comes from the dynamics of the system and not 
from random location of boundary points. 
Thus we have located the boundary points in a regular way, with perimeter spacing 
\be 
\fl d(x_{n+1},x_{n}) = \frac{d}{N-1}, ~  n=1,\dots,N-1 ~ .
\ee
The rule of dynamics is free motion in the 
interior region and elastic specular collision at the central 
disc and the exterior square wall. 
%
We have classified trajectories using the scheme presented in section \ref{sec:matrix_construction}. 
A global characteristic feature of chaos is encoded in the number of classical trajectories. 
For the Sinai billiard we found that the number $N_{\mathrm{traj}}$ of trajectories 
averaged over all pairs of boundary points increases as function of the number of 
rebounds $N_{\mathrm{reb}}$ like an exponential, 
\be
\label{eq:SinaiSolReb}\fl
N_{\mathrm{traj}} = \exp(\alpha ~ N_{\mathrm{reb}}) - 1 ~ , ~ \alpha= 0.81 \pm 0.03 ~ .
\ee
%
Such exponential behaviour in chaotic billiards is 
clearly distinct from the behaviour in integrable billiards, where the 
number of trajectories increases linearly with the number of rebounds 
(see rectangular billiard). We fixed a value for $N_{\mathrm{reb}}$ and then generated an ensemble of length matrices 
$\{\Lambda^{\nu}|\nu=1,\dots,N_{\mathrm{traj}}\}$ corresponding to all possible 
trajectory indices $\nu$. In general one expects that chaotic behaviour develops 
with increasing $N_{\mathrm{reb}}$.

In order to make sure that the chaotic behaviour is not due to numerical noise, we
estimated the numerical error $|x_{fi}-x_{in}|$ by following a trajectory from 
a given starting point $x_{in}$ until it carried out $N_{\mathrm{reb}}$ rebounds, 
then following the trajectory in time reversed direction to arrive after 
another $N_{\mathrm{reb}}$ at some $x_{fi}$. We observed that the error behaviour has two regimes. 
For $N_{\mathrm{reb}} < 48$ it shows 
an exponential increase, following on average the rule 
\be
\label{eq:SinaiErrReb}\fl
\epsilon = \exp(\beta ~ N_{\mathrm{reb}} - 41) ~ , ~ \beta = 0.76 \pm 0.03 ~~ \mbox{for} ~ N_{\mathrm{reb}} < 48.
\ee
For $N_{\mathrm{reb}} > 48$ saturation is reached (the relative error is in the order of one). 
Such exponential behaviour is related to the largest positive Lyapunov 
exponent of the system and thus represents a global characteristic of a chaotic system.
Like the exponential behaviour of number of trajectories, equation~(\ref{eq:SinaiSolReb}), 
also the exponential error behaviour distinguishes between chaotic billiards 
and integrable billiards.

We found a relative error of about  
$10^{-13}$ for $N_{\mathrm{reb}} \le 15$.
After unfolding the spectra we retain 8 significant digits. 
We used the technique of Gaussian broadening~\cite{Haake01,Bruus97} to unfold 
the raw spectrum. In this method there is a free parameter, 
which we tuned to reproduce the auto-correlation 
coefficient ($C=-0.27$ for Wigner distribution~\cite{Bohigas83,Seligman84}). 
For a given trajectory index $\nu$ we computed a level spacing distribution. Afterwards 
we superimposed the spectra corresponding to different trajectory 
indices $\nu$. For $N_{\mathrm{reb}}=15$ the number of trajectories is quite 
large ($N_{\mathrm{traj}} \sim 10^{5}$). In order to limit computational cost, 
we have restricted ourselves by considering smaller subsets 
of length matrices. 
%
It turns out that $n_{\mathrm{traj}}=20$ gives a reasonably good statistics. Note that, in this way, we don't consider all trajectories and have to order them arbitrarily as explained in section \ref{sec:matrix_construction}.
Therefore, we verified that the resulting distributions are 
independent of the particular subset of trajectories and ordering used within statistical 
errors. The resulting level spacing distribution $P(s)$ of length eigenvalues 
and the spectral rigidity $\Delta_{3}(L)$, corresponding to $N=50$  boundary points, 
$N_{\mathrm{reb}}=15$ rebounds and $n_{\mathrm{traj}}=20$ trajectories, are shown in figure \ref{fig:SinaiLength+LSD+Delta3}(b-c).
The results show a Wigner distribution for $P(s)$, and are consistent with GOE 
behaviour also for $\Delta_{3}(L)$. As in the quantum Sinai billiard, the level 
spacing distribution has been shown to give a Wigner distribution~\cite{Bohigas84}. 

The histogram of the length matrix elements itself is
shown in figure \ref{fig:SinaiLength+LSD+Delta3_top}. The distribution 
looks close to a Gaussian. 
Determining if it is a pure Gaussian, is a question physically relevant 
for the following reason: If the distribution $P(\Lambda)$ is a Gaussian, 
then the matrix elements obey GOE statistics. Then RMT~\cite{Mehta91} 
implies that the level fluctuation statistics $P(s)$  
and the spectral rigidity $\Delta_{3}(L)$ follow GOE statistics 
(which seems consistent with the numerical results shown in 
figure \ref{fig:SinaiLength+LSD+Delta3}(b-c). 
We pointed out above that the random walk model 
gives a Gaussian distribution for the histogram of the length matrix 
elements $P(\Lambda)$. The random walk model is mathematically a 
Markov chain, i.e., the length of each piece of straight path in 
between subsequent collisions is given by a random number. Two 
subsequent random numbers are statistically independent. On the other 
hand, the chaotic Sinai billiard is a deterministic system, and the 
length of two subsequent pieces of straight trajectory are \emph{not} independent. 
Hence it is plausible that the distribution $P(\Lambda)$ of length matrix 
elements for the Sinai billiard is not given by an exact Gaussian. 

{\it Mathematical note.} The BGS-conjecture does {\it not} 
state that the matrix elements of a quantum Hamiltonian must be 
distributed like a GOE ensemble. The conjecture rather only says 
that the statistical fluctuations of the eigenvalue spacings 
obtained from the quantum Hamiltonian are the same 
as those from a GOE ensemble, giving a Wignerian distribution. In other words, 
it is possible that the matrix elements of the quantum Hamiltonian 
be distributed quite differently from a Gaussian and that 
its level spacing distribution be nevertheless Wignerian. 

Such a situation, where the distribution of matrix elements is not GOE, 
but the level fluctuation statistics is GOE, occurs in nuclear physics.
An example is the distribution of the Hamiltonian matrix elements 
obtained from nuclear shell model calculations~\cite{Guhr98}. 
In this model, there are vanishing Hamiltonian matrix elements. 
This implies that the number of independent matrix 
elements is much smaller than in a random matrix of the same size. 
However, Brody et al.~\cite{Brody81} showed that the 2-body 
residual interaction in the shell model yields matrix elements of 
random character following a Gaussian distribution. In particular, 
they showed that spectral fluctuation properties from such ensembles 
with orthogonal symmetry are identical to those from GOE. That implies 
that GOE is meaningful to predict spectral fluctuation properties 
of nuclei governed by 2-body interactions, though the Hamiltanian 
does not follow a Gaussian distribution.

\subsection{Bunimovich Stadium Billiard}
\begin{figure}[ht]
\centering
	\label{fig:StadiumGeometry}
	\includegraphics[width=5.5cm]{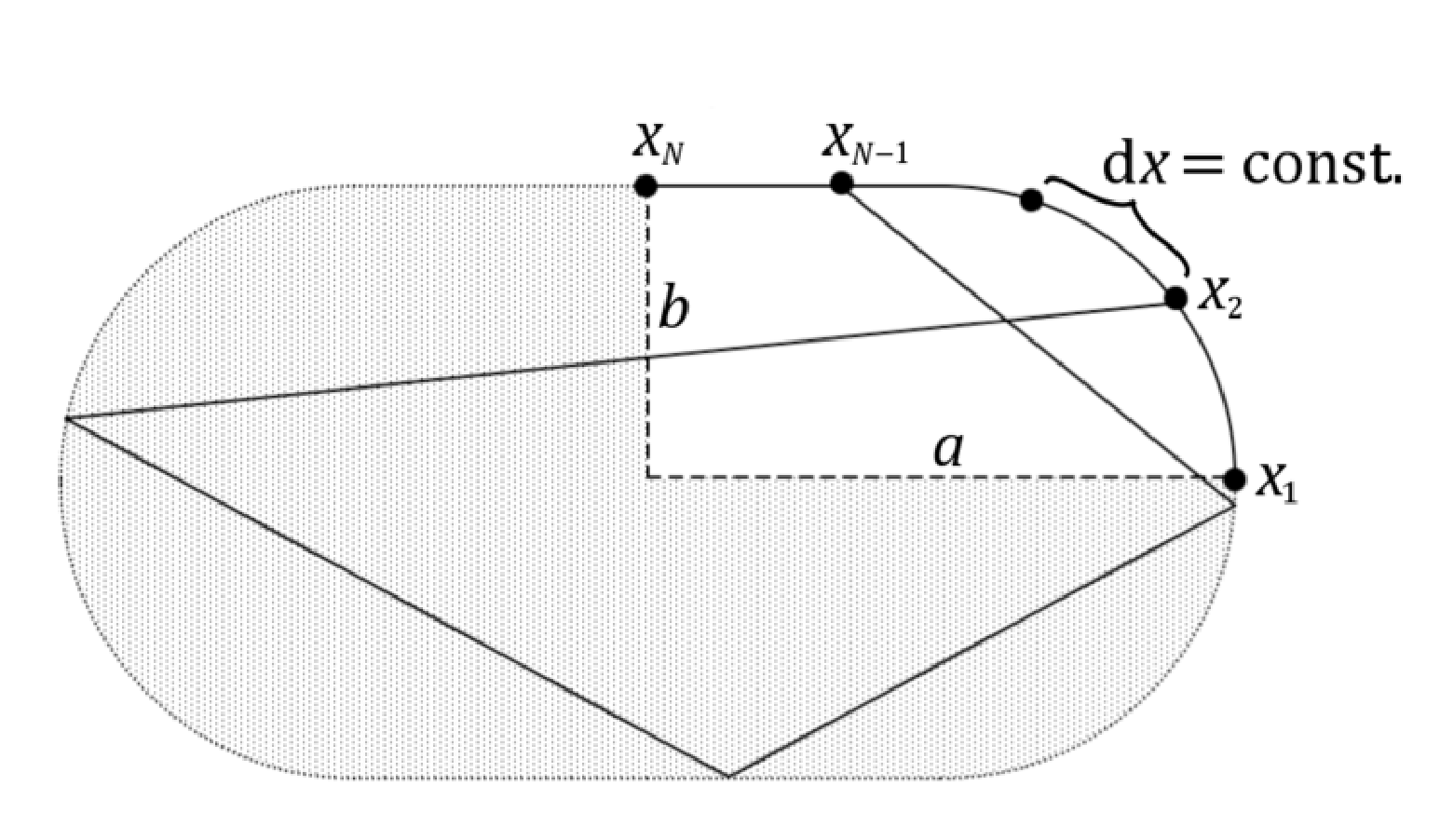}
\caption{2D stadium billiard. Trajectories go from boundary points $x_{i}$ to $x_{j}$ located in a segment of one quarter of the billiard wall.
}
\end{figure}
%
%
%
%
%
%
%
%
\begin{figure}[ht]
\centering
\subfigure[]{\label{fig:StadiumAverTrajLength+LSD+Delta3_top}
\includegraphics[width=6cm]{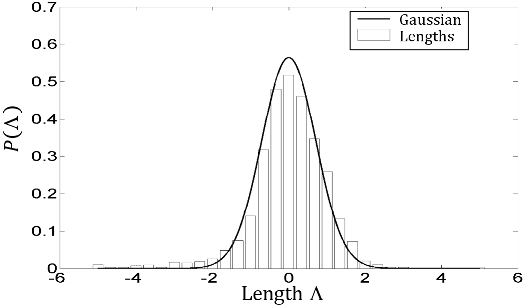}}\\
\subfigure[]{\label{fig:StadiumAverTrajLength+LSD+Delta3_middle}
\includegraphics[width=6cm]{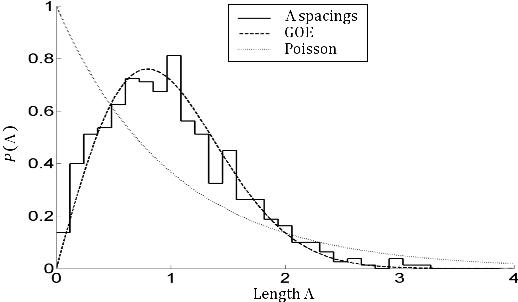}}\\
\subfigure[]{\label{fig:StadiumAverTrajLength+LSD+Delta3_bottom}
\includegraphics[width=6cm]{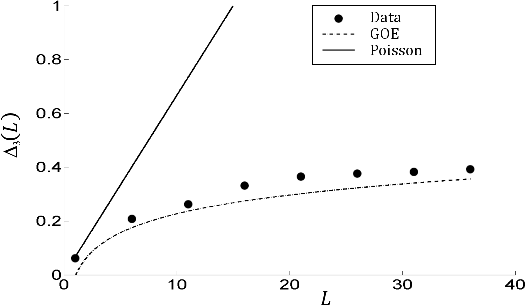}}
\caption{2D stadium billiard.  Average over trajectories. Geometry parameters $a/b=2.2$. Boundary points on quarter of wall $N=40$.
(a) Distribution of length matrix elements (after subtraction of mean $\langle\Lambda\rangle$). $N_{\mathrm{reb}}=30$.  
(b) Level spacing distribution $P(s)$ from length matrix $\Lambda$. $N_{\mathrm{reb}}=12$. Average over $n_{\mathrm{traj}}=18$ trajectories.
(c) Spectral rigidity $\Delta_{3}(L)$, same parameters as (b).}
\label{fig:StadiumAverTrajLength+LSD+Delta3}
\end{figure}
Let us consider the 2D Bunimovich stadium billiard with semi-axes $a$ and $b$ 
(figure \ref{fig:StadiumGeometry}). The billiard is known 
to be fully chaotic~\cite{Bunimovich74}. The billiard is symmetric 
under mirror operation about $x$- and $y$-axes. In order not to mix different 
symmetry classes, we consider the billiard with boundary points 
$x_{1},\dots,x_{N}$ located in one quarter of the billiard wall. 
The quarter perimeter has the length $b+\pi(b-a)/2$. 
The nodes $x_{n}$ are equidistantly distributed on the quarter perimeter 
of the stadium, with perimeter spacing given by 
\be\fl
d(x_{n+1},x_{n})=\frac{[b+\pi(b-a)/2]}{(N-1)}\,,\;n=1,\dots,N-1.
\ee
The rule of dynamics is free motion in the interior region and  
elastic specular collision at the exterior square wall. 
For a given pair of boundary points, the number of trajectories $N_{\mathrm{traj}}$ 
connecting these points increases on average with the number of rebounds 
exponentially, according to 
\be
\label{eq:StadiumSolReb}\fl
N_{\mathrm{traj}} = \exp[\alpha ~ (N_{\mathrm{reb}} - 1)] ~ , ~ \alpha=1.01 \pm 0.09 ~ .
\ee
%
Such behaviour is qualitatively similar 
to that found in the Sinai billiard. We also measured the numerical error 
following the method used in the Sinai billiard. Likewise, we found a regime of exponential 
behaviour followed by a regime of saturation (not shown).
On average the exponential increase is given by 
\be
\label{eq:StadiumErrReb}\fl
\epsilon = \exp(\beta ~ N_{\mathrm{reb}} - 14.3)\,,\;\beta=0.247 \pm 0.005,\;N_{\mathrm{reb}} < 58.
\ee
We have classified trajectories by the the number of rebounds 
$N_{\mathrm{reb}}$, which was kept fixed. 
In this way, we generate an ensemble of length matrices 
$\{\Sigma^{\nu}|\nu=1,\dots,N_{\mathrm{traj}}\}$ corresponding to all possible 
trajectory indices $\nu$. For each trajectory index $\nu$ we computed 
a level spacing distribution.
Afterwards we superimposed the spectra coming from a smaller subset of trajectory indices 
$\nu=1,\dots, n_{\mathrm{traj}}$ (like we did for the Sinai billiard). 
The resulting level spacing distribution $P(s)$ 
of length eigenvalues and the spectral rigidity $\Delta_{3}(L)$ are shown in 
figure \ref{fig:StadiumAverTrajLength+LSD+Delta3}(b--c). 
The results are consistent with a Wigner distribution, i.e. GOE behaviour. 
The histogram of the length matrix elements $P(\Lambda)$ shown in 
figure \ref{fig:StadiumAverTrajLength+LSD+Delta3_top} looks close to  
a Gaussian.

\begin{figure}[ht]
\centering
\subfigure[]{\label{fig:StadiumAverShapeLength+LSD+Delta3_top}
\includegraphics[width=6cm]{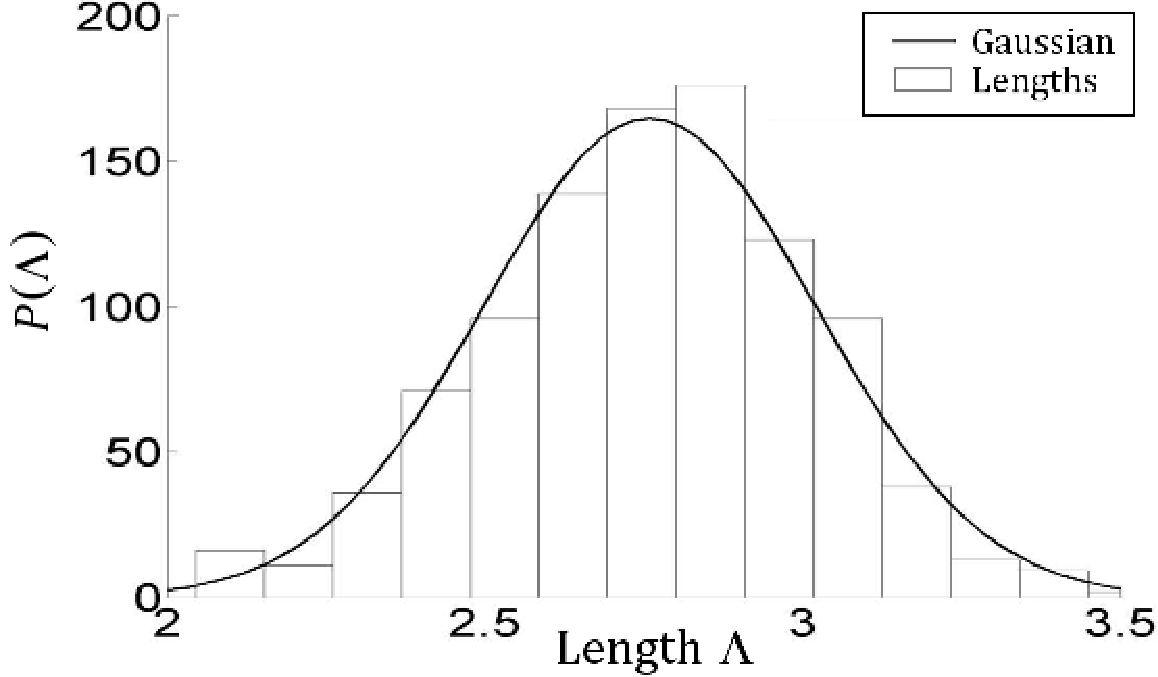}}\\
\subfigure[]{\label{fig:StadiumAverShapeLength+LSD+Delta3_middle}
\includegraphics[width=6cm]{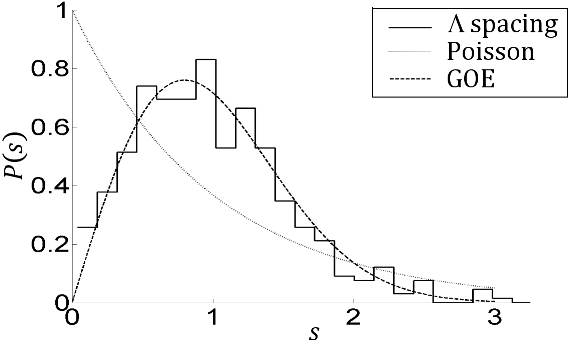}}\\
\subfigure[]{\label{fig:StadiumAverShapeLength+LSD+Delta3_bottom}
\includegraphics[width=6cm]{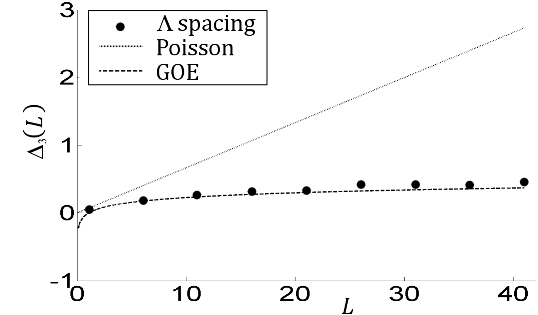}}
\caption{2D stadium billiard. Average over shapes of stadium. 
(a) Distribution of length matrix elements. $N_{\mathrm{reb}}=12$.
(b) Level spacing distribution from length matrix $\Lambda$. 
(c) Spectral rigidity $\Delta_{3}$. }
\label{fig:StadiumAverShapeLength+LSD+Delta3}
\end{figure}

In order to see if the observed Wigner distribution in the level 
spacing distribution depends on the statistical method of averaging 
over several trajectories (i.e. length matrices $\Lambda^{\nu}$) we 
have tested an alternative, where we averaged over different stadium 
geometries. All stadium billiards are fully chaotic, and if the Wigner 
behaviour is universal, this should show up for any stadium geometry. 
Superimposing spectra corresponding to different stadium shapes should 
improve statistics. In their paper on quantum chaos in nuclear physics, 
Bohigas et al.~\cite{Bohigas83b} have analyzed spectra of a series 
of different heavy nuclei (but having the same quantum numbers) and 
showed that their level fluctuation statistics agrees with the prediction 
of RMT. These different nuclei correspond to 
different potentials. In analogy to that we considered here length 
spectra corresponding to different stadium shapes. The stadium wall 
represents a curve where the potential jumps from zero to infinity. 
Hence different stadium shapes correspond to different potentials. 
By superimposing spectra from different stadium shapes we obtained 
results displayed in figure \ref{fig:StadiumAverShapeLength+LSD+Delta3}. 
Within statistical errors the results are equivalent to those obtained 
by superimposing spectra from different trajectories 
(figure \ref{fig:StadiumAverTrajLength+LSD+Delta3}).

\section{Universality in chaotic billiards} 
\label{sec:universality}
The numerical experiments with chaotic billiards investigated above 
led us to the following observations:
(i) For a given pair of boundary points 
$N_{\mathrm{reb}}$, the number of classical trajectories $N_{\mathrm{traj}}$ shows exponential 
behaviour as function of $N_{\mathrm{reb}}$. 
(ii) For a given pair of boundary points, 
the numerical error $\epsilon$ shows exponential behaviour as function of the 
number of rebounds $N_{\mathrm{reb}}$. 
(iii) The NNS level fluctuation distribution $P(s)$ 
obtained from eigenvalues of the length matrix $\Lambda$ shows a 
Wignerian distribution. This is consistent with GOE behaviour predicted by RMT.
(iv) The spectral rigidity $\Delta_{3}(L)$ obtained from the spectrum of length 
matrix $\Lambda$ is also found consistent with GOE behaviour predicted by RMT.
(v) The distribution $P(\Lambda)$ of length 
matrix elements $\Lambda_{ij}$ is found to be close to a Gaussian. 

The leading Gaussian behaviour in the distribtion $P(\Lambda)$ is possibly universal. 
How can we understand such behaviour of $P(\Lambda)$ in chaotic billiards?
The Sinai billiard is equivalent 
to a system where the billiard ball moves in an open system of 
equal spherical discs located on a 2D rectangular regular grid, known as 
the 2D (diluted) Lorentz gas model. The billiard ball alternates between free motion and 
collisions with the circular discs. There are two types of 2D periodic Lorentz gases 
(or Sinai billiards on a torus): One has a finite horizon, where free paths between collisions are 
bounded (the scatterers are sufficiently dense to block every direction of motion. 
The other type has an infinite horizon, where the 
particle can move indefinitely without collision with any disc. The Sinai billiard investigated 
above (see figure \ref{fig:SinaiGeometry}) belongs to this class. 
The CLT, the existence of finite diffusion coefficient and the convergence to Brownian motion 
were proven only for the periodic Lorentz gas with finite horizon~\cite{Bunimovich81,Bunimovich91}.
In the case of infinite horizon, the CLT has not been proven. Moreover, in this case the diffusion 
coefficient is infinite and there is no convergence to Brownian motion~\cite{Szasz07}.

Actually, for the periodic Lorentz gas with finite horizon it can be shown rigorously that the 
distribution of length of trajectories becomes a Gaussian distribution 
in the limit of many bounces. This holds when the initial points of trajectories 
are distributed randomly on the billiard boundary.
Then Chernov and Markarian~\cite{Chernov06a} prove 
the following result 
\be\fl
\lim_{n \to \infty} \mu \left( \frac{ t_{n} - n\langle \tau \rangle }{ \sqrt{n} } \le z \right) = \frac{1}{ \sqrt{2 \pi} \sigma_{f} }
\int_{-\infty}^{z} \rme^{-s^{2} / 2 \sigma^{2}_{f} }\;\rmd s
\ee
for all $-\infty < z < \infty$. Here $\tau(x)$ denotes the return function and
\be
\fl t_{n} = \tau(x) + \tau(F x) + \cdots + \tau(F^{n-1} x) 
\ee
is the time of the n-th collision. $\sigma^{2}_{f}$ is the variance of the random 
variable depending on the observable $f$, which here is the return function 
$\tau(x)$. $F:M \to M$ is the collision map and $M$ is the collision space 
(phase space of billiard map). Thus the distribution of the travel time until 
the $n$-th collision obeys the CLT, and hence converges in distribution to a normal, 
i.e. Gaussian distribution in the limit where the number of collisions goes to infinity.
In the case where the billiard particle moves with constant speed $u$, 
this means also the length of trajectory $\Lambda_{n} = u t_{n}$ obeys the CLT, 
and the length distribution $P(\Lambda_{n})$ tends to a normal Gaussian distribution for 
$n \to \infty$. 

This result seems to support our numerical findings of a (near) Gaussian 
distribution of length of trajectories. However, the scenario where the 
above mathematical result holds and the scenario of our numerical study differ 
in two respects, namely in the distribution of boundary points and in the 
horizon of billiard. For the purpose of statistical analysis in terms of 
RMT we are interested in the distribution $P(\Lambda_{n})$, which corresponds 
to the case where the billiard particle starts from 
and arrives at boundary positions taken from a {\it discrete} set of boundary points, 
which is different from an initial random set of boundary points.  
Nevertheless, it seems plausible that a similar mathematical result may hold 
also for the discrete set of initial and final boundary points, such that 
$P(\Lambda_{n})$ tends to a Gaussian. Second, the mathematical result holds 
for the case of finite horizon, while the Sinai billiard in our numerical study 
has an infinite horizon. At a first glance, it looks puzzling in absence of 
strong mathematical results (for infinite horizon) that our 
numerical results nevertheless show (near) Gaussian behaviour for the length 
distribution. A possible explanation is this: The choice of the discrete 
set of regularly distributed boundary points selects preferentially 
trajectories which often bounce at the central disc, while trajectories 
with infinite horizon (no bounces at disc) are avoided. 
This is supported also by numerical evidence that the distribution 
$P(\Lambda_{n})$ of the Sinai billiard (figure \ref{fig:SinaiLength+LSD+Delta3_top} 
and that of the rectangular billiard viewed as Sinai billiard with central disc of radius 
zero (figure \ref{fig:RectangleStatistics_top} 
are quite different. \\
\indent In the case of the stadium billiard, B\'{a}lint and Gou\"{e}zel~\cite{Balint06} have shown 
that a limit theorem also holds. They proved that the limit distribution ($n \to \infty$) 
is Gaussian, however, the scaling factor is not given by the standard expression 
$\sqrt{n}$, but rather by $\sqrt{n \log n}$. As consequence, there 
is no finite standard diffusion coefficient and no convergence to Brownian motion. 
B\'{a}lint and Gou\"{e}zel prove that the distribution of length of trajectories, 
tends to a gaussian distribution in the limit where the number of collisions 
$n$ goes to infinity. Like in the Lorentz gas, this corresponds to a random 
distribution of initial points on the boundary. This is consistent with
our numerical results of the stadium billiard which show in $P(\Lambda)$ small 
deviations from a Gaussian comparable in magnitude with those in the Sinai 
model. They are likely due to the small number of bounces. \\
\indent Let us suppose that $P(\Lambda_{n})$ does tend to a Gaussian. 
As a consequence, in this limit (and after suitable normalization) 
the length matrix $\Lambda$ itself will become a GOE matrix~\cite{Stockmann99}. 
This is sufficient to guarantee in this limit that the level fluctuations $P(s)$ 
and spectral rigidity $\Delta_{3}(L)$ obtained from the matrix $\Lambda$ show 
GOE behaviour~\cite{Mehta91}. This does establish universal behaviour 
in the limit of many bounces in the case of the rectangular Lorentz gas/Sinai billiard 
as well as the Bunimovich stadium billiard.  \\
\indent Now we want to address the following questions: \\
(1) Concerning universality observed in classical chaotic billiards,
what are the underlying physical principles? 
We will give a heuristic description---not a rigorous derivation---of the 
physical principles leading to the phenomenon of universality. Let us consider 
chaotic billiards in the regime of macroscopic times, i.e., when the billiard particle does 
a large number of bounces.
Consider a trajectory carrying out $N_{\mathrm{reb}}$ bounces. Because 
the system is a classically deterministic system, each segment of trajectory 
between two subsequent bounces depends on and is completely determined by the 
preceeding segment of trajectory. However, if one computes the correlator between 
trajectory segment $n$ and trajectory segment $n+k$, such correlation tends to zero, 
when $k$ becomes large (see below equations (\ref{eq:CorrExpFallOff}) and (\ref{eq:CorrPolynomFallOff}),
and reference \cite{Chernov06a}).
This effect is due to the dynamics of classical chaos (positive Lyapunov exponent).
Thus two segments of the same trajectory, which are sufficiently distant in terms 
of travel time or number of intermediate bounces, will become statistically independent.
This makes the system similar to a random walk system, where any two segments of a
trajectory are statistically independent. To see this, consider trajectories of the 
chaotic billiard composed of $N$ segments ($N-1$ bounces), and choose $N \gg k$. 
Then we can consider all segments $i, j$ with $|i-j| > k$ to be statistically independent.
Consequently, for $N \gg k$ almost all segments are statistically independent 
from each other and hence the system will statistically behave similar to a random walk system. 
This means in the regime of a large number of rebounds that the system looses its 
memory of the detailed underlying laws of physics. It becomes universal. 
Such universality manifests itself in the statistical observable of distribution of length of 
trajecories $P(\Lambda)$ which for all chaotic billiards studied here shows a near-Gaussian 
character (like the random walk system). \\
\indent (2) What is the physical significance of such universality? 
If one considers classically chaotic billiards in the regime of macroscopic times, 
where $\Lambda$ denotes the matrix of length of trajectories, and if one considers as statistical observable the
distribution of length of trajectories $P(\Lambda)$, then $P(\Lambda)$ will asymptotically 
approach a Gaussian and $\Lambda$ will asymptotically become a random matrix from a 
Gaussian Orthogonal Ensemble (GOE). Hence the level spacing distribution of the length matrix $\Lambda$ 
will be described by GOE random matrix statistics.
However, one should be careful. The matrix of length $\Lambda$ has not 
lost all information on the physics of the underlying system. For instance, 
the variance of length $\Delta \Lambda^{2}$ turns out to behave differently 
in the Sinai billiard (on a torus) in the finite horizon regime, scaling like $n$, 
in contrast to the stadium billiard, where it  scales like $n~\log(n)$. Statistical behaviour depends on the 
particular observable considered. Thus using the length marix $\Lambda$, which 
via $P(\Lambda)$ and $P(s)$ displays universal behaviour,
we can compute physical observables, e.g., being related to $\Delta \Lambda^{2}$. 
Such observable yields transport properties, in particular the diffusion coefficient. 
Results of transport properties are presented in section \ref{Sec-Transport}.\\
\indent (3) Are there connections between universality and physical quantities 
which are easily observed in real systems?
In particular, is such universality related to transport properties of the classical system? 
The answer is yes, and we will show in section \ref{Sec-Transport} how a transport coefficient can be obtained 
from the length matrix $\Lambda$ in the regime of universality. \\
\indent Moreover, the behaviour  
of the chaotic billiard systems when approaching the regime of universality is 
characterized by laws specific for the particular billiard system.
As example consider the decay laws of correlation functions. 
Consider a billiard trajectory and consider as observable $f$ the  segment of trajectory 
from $n$ to $n+1$, while $g$ denotes the segment from $n+k$ to $n+1+k$. 
Then the correlation function between observables $f$ and $g$ for many billiards shows 
either an exponential fall-off behaviour \cite{Chernov06a} 
\be
\label{eq:CorrExpFallOff}
|C_{f,g}(k)| \le Ae^{-\alpha |k|} ~ ,
\ee
or polynomial fall off behaviour
\be
\label{eq:CorrPolynomFallOff}
|C_{f,g}(k)| \le B|k|^{-\alpha} ~ ,
\ee
where $A$ and $B$ are constants, $k\ne 0$ and the exponent $\alpha>0$ is specific for the particular billiard.
For example, the triangular Lorentz gas as well as the Bunimovich stadium billiard 
obey exponential falloff behaviour \cite{Chernov06a}. \\
 \indent Here we suggest for chaotic billiard systems that the approach towards universality 
(i.e., increasing the number of rebounds) contains further physical information 
characteristic for the system. In particular, we expect that the distribution $P(\Lambda_{N_{reb}})$ 
approaches a Gaussian,
\be
P(\Lambda_{N_{reb}}) \to P_{Gaussian},
\ee
for $N_{reb} \to \infty$, where the asymptotic approach is characterized by an exponent 
characteristic for the particular billiard. \\
\indent (4) Are those universality properties related to thermodynamical observables? 
If we consider the chaotic billiards consisting of  a single particle moving in a rigid environment of scatterers, 
then it does not make sense to talk about thermodynamics. Thus for systems considered in this work the answer is no.
However, if one considers billiards of many particles, then thermodynamics (as a function of temperature) 
will influence the dynamics. We shall defer the study of such effects on universality to future work. \\
\indent (5) Is such universality related to spectral statistics of the corresponding quantum system, i.e., 
what is the relation between universality in classical chaotic systems 
and universality in chaotic quantum systems, as defined via the 
Bohigas-Giannoni-Schmit conjecture \cite{Bohigas84}? 
This is a very interesting question, for which we do not  
have an answer. The finding that universality properties exist in both, 
classical and quantum chaos, may be a hint that such a relation actually exists.
On the other hand, for universality in quantum chaos in the semi-classical regime, 
periodic orbits play a crucial role (via Gutzwillers trace formula). In contrast, 
in our study of chaotic classical billiards universality is captured in the 
length matrix of bouncing, zig-zag, non-periodic trajectories. 
In order to address such question further, we suggest to consider the path 
integral relation (in Euclidean, i.e. imaginary time)
\be
\fl \langle y|\exp[-HT/\hbar]|x\rangle = \int [\rmd x]\left.\exp[-S_{\mathrm{Eucl}}[x]\right|_{x,0}^{y,T} ~ ,
\ee
which relates the quantum Hamltonian $H$ to the classical Euclidean action 
$S_{\mathrm{Eucl}}$. For billards, recall that the length of a trajctory is essentially 
equivalent to its action. Thus spectral properties of the quantum Hamiltonian 
may indeed be related to the spectrum of the classical action.
%
%
%
%
%
\begin{figure}[ht]
\centering
\subfigure[]{\label{fig:StadiumDiffusion_top}
\includegraphics[width=6cm]{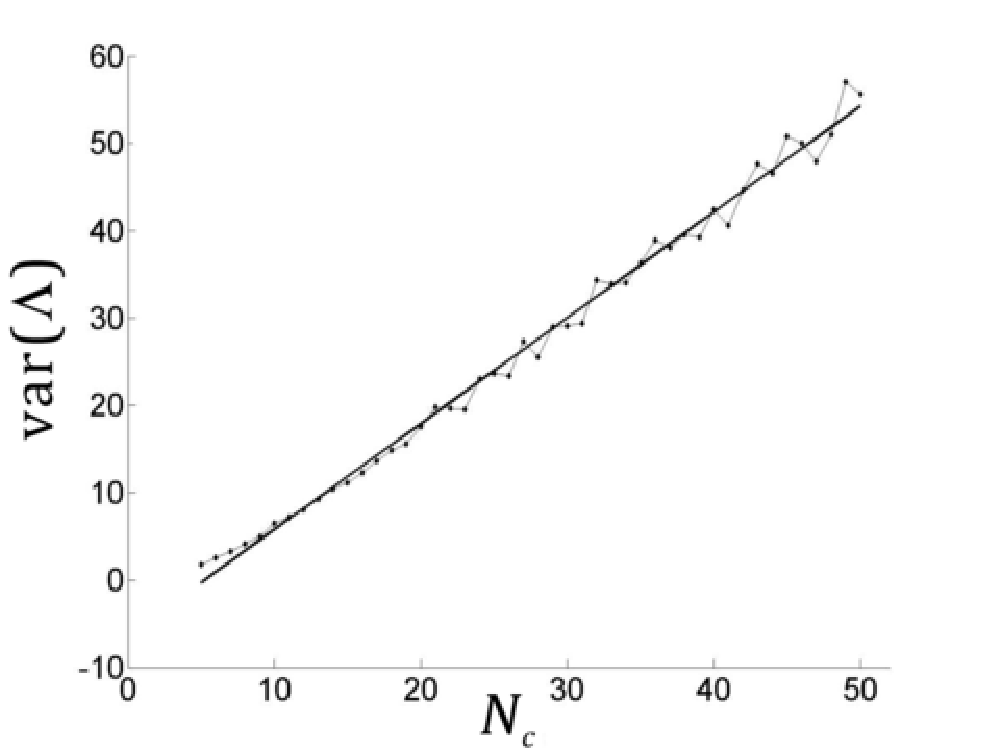}}\\
\subfigure[]{\label{fig:StadiumDiffusion_middle}
\includegraphics[width=6cm]{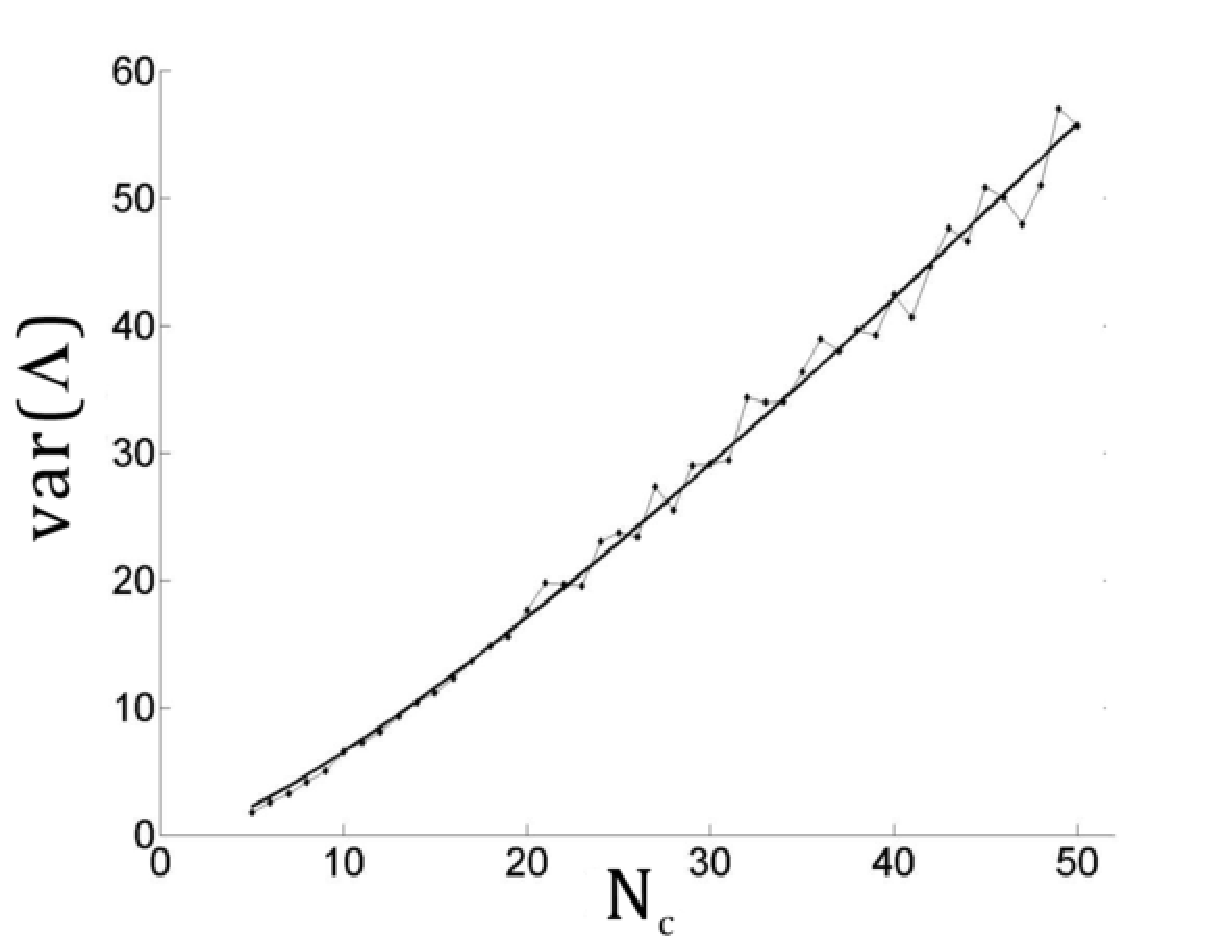}}\\ 
\subfigure[]{\label{fig:StadiumDiffusion_bottom}
\includegraphics[width=6cm]{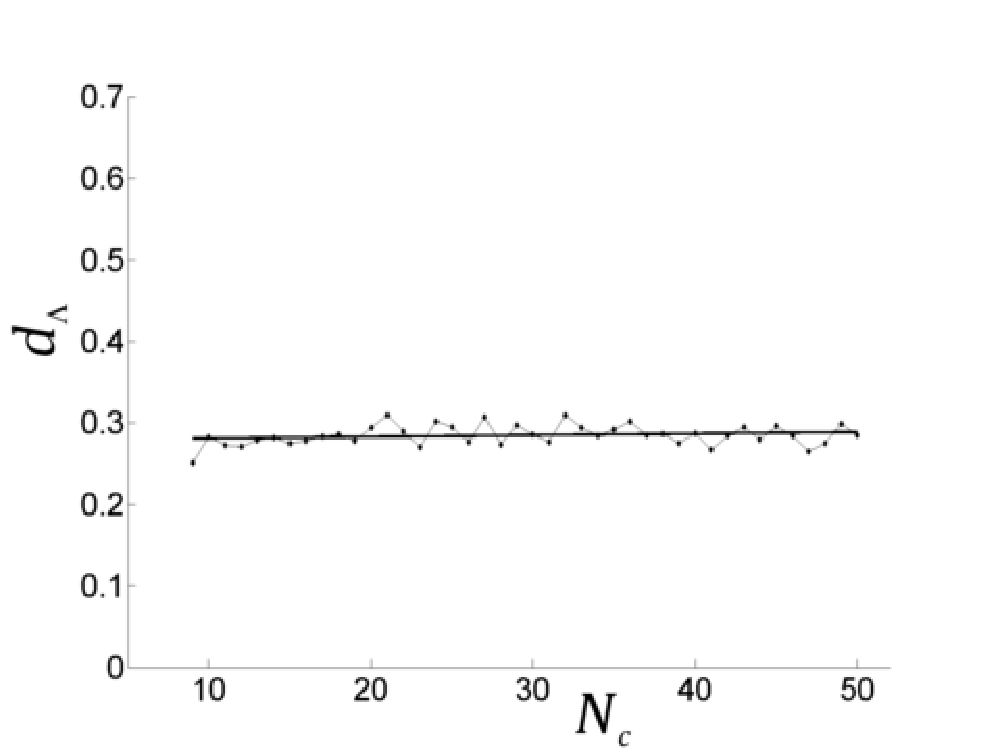}}
\caption{Anomalous diffusion in stadium billiard. 
(a) Variance of $\Lambda$ linear fit in $n$. 
(b) Variance of $\Lambda$ fit including logarithmic term  $n log(n)$. 
(c) Diffusion constant from including logarithmic term.}
\label{fig:StadiumDiffusion}
\end{figure}

\section{Transport properties from length of trajectories}
\label{Sec-Transport}
Above we have shown for the Bunimovich stadium billiard and for the Sinai 
billiard that they display universality properties via the statistical behaviour 
of the matrix of length of trajectories. Here we will show that such universal behaviour 
is related to relevant physical quantities. In particular we will extract transport 
properties from the length matrix (note the analogy to computation of transport 
properties in quantum chaos in semi-classical regime via Gutzwiller formula).
As examples we consider the 
stadium billiard.  



In systems for which the CLT is verified, the diffusive character is manifested by a linear relation between the time of travel 
and the variance of position. The diffusion coefficient $d$ (in 2 dimensions) is given by 
the Einstein relation
\be
d = \frac{1}{4  \tau} ~ \mathrm{var}( \vec{X} )
\ee
where $\tau$ denotes the time of travel and $\mathrm{var}( \vec{X} )$ denotes the variance of 
the position. In~\cite{Hosseinizadeh11}, a similar linear relation between the time of travel and 
the variance of the length of trajecories is demonstrated 
for another system verifying the CLT, the periodic triangular Lorentz gas with finite horizon:
\be
	\mathrm{var}( \Lambda ) \propto \tau ~ ,
\ee
where travel time $\tau$ is related to mean trajectory length of $n$ collisions  $\langle\Lambda_{n}\rangle$ 
and velocity $u$ via $\tau=\langle\Lambda_{n}\rangle/u$.

Based on this approach, we define a diffusion coefficient with respect to the variable $\Lambda$:
\be
	\label{eq:dLambda}
	d_{\Lambda} = \frac{\mathrm{var}( \Lambda )}{n} ~ . 
\ee
for the Bunmovich stadium billiard, which is a chaotic system 
with concave repeller/scatterer. We have chosen to analyze such system, and compute its 
transport properties because it reveals a very interesting non-standard diffusion behaviour.
In a numerical modelling study of the stadium billiard, Borgognoni et al. \cite{Borgonovi96}
have studied its diffusion behaviour by testing if a linear relation holds between the 
time of travel and the variance of angular momentum. They found that a linear relation fits 
quite well their data and computed the diffusion coefficients as ratio of variance of 
angular momentum and time of travel (or number of bounces). 
However, in 2006, B\'{a}lint and Gou\"{e}zel \cite{Balint06} proved rigorously for the 
stadium billiard that a \textquotedblleft non-standard\textquotedblright\ limit theorem holds, where the width does 
not follow the usual $\sqrt{n}$ as in equation (\ref{eq:dLambda}), but rather follows an anomal $\sqrt{n \log n}$ law. 
As consequence, the system does not converge to Brownian motion in the limit of many 
bounces, and the standard diffusion coefficient does not exist.

We carried out numerical simulations using trajectories of the length matrix $\Lambda$ 
and calculated the total time of travel $\tau$ and the variance of length 
$\mathrm{var}( \Lambda )$. According to reference \cite{Balint06}, one expects that the 
variance should scale like the travel time with a logarithmic correction 
\be
\mathrm{var}( \Lambda ) \propto \tau ~ \log(\tau) ~ .
\ee
We define the diffusion coefficient by
\be
d_{\Lambda} = \frac{\mathrm{var}( \Lambda )}{n ~ \log(n)} 
\ee
We have done statistical tests of $\Delta \Lambda^{2}$ comparing a fit linear in $n$ 
with a fit linear in $n ~ log(n)$. The results are shown in figure \ref{fig:StadiumDiffusion}. 
Making fits of the form
\bea
\mathrm{var}( \Lambda ) &=& a_{1} + b_{1} ~ n ~  ,
\nonumber \\
\mathrm{var}( \Lambda ) &=& a_{2} + b_{2} ~ n \log(n) ~ ,
\eea
we obtained for the linear fit $a_{1} =-6.32, b_{1} = 1.21$ with an error of 
$2.8$ (max) and $7.2 \times 10^{-2}$ (mean)
shown in figure \ref{fig:StadiumDiffusion_top},
compared to the fit with logarithmic correction  giving $a_{2} =-0.0831, b_{1} = 0.286$ with an error of 
$4.1 \times 10^{-2}$ (max) and $3.8 \times 10^{-3}$ (mean)
shown in figure \ref{fig:StadiumDiffusion_middle}.
One observes that the statistical error of the fit including the logarithmic correction 
is smaller than that of the linear fit.
An estimate of the anomalous diffusion constant from the logarithmic fit is shown in 
figure \ref{fig:StadiumDiffusion_bottom}.
This means (i) numerical consistance with the mathematical result obtained by B\'{a}lint and 
Gou\"{e}zel \cite{Balint06}, (ii) the Bunimovich stadium billiard, although considered a 
non-generic billiard in quantum chaos due to the \textquotedblleft bouncing ball states\textquotedblright\ of wave solutions, 
is a very interesting system in classical chaos in the asymptotic regime of many bounces, 
which displays anomalous diffusion behaviour, (iii) the length matrix $\Lambda$ which 
generates universal behaviour in $P(\Lambda)$ and $P(s)$, contains information allowing 
to distinguish normal diffusion from anomalous diffusion.

\section{Summary}
This paper is about classical chaos occuring widely 
in nature, for example in astrophysics, meteorology and dynamics 
of the atmosphere, fluid and ocean dynamics, climate change, 
chemical reactions, biology, physiology, neuroscience, or medicine.  
We have suggested to extend random matrix theory, used in chaotic 
quantum systems, to classically chaotic and integrable systems. 
We have studied fully chaotic as well as integrable billiards and used 
a statistical description based on the length of trajectories 
to discriminate chaotic versus integrable behaviour. 
\\
{\it Results:}\\ 
(i) In chaotic billiards (stadium and Sinai billiard) 
the NNS distribution $P(s)$ as well as $\Delta_{3}(L)$ obtained 
from the length matrix show GOE behaviour, i.e., they are universal. 
The implication is that RMT not only models 
spectral statistical fluctuation properties in chaotic quantum systems, 
but goes beyond and applies as well to chaotic classical systems.
\\
(ii) The distribution of length matrix elements $P(\Lambda)$ itself 
is universal to leading order. The difference between the shape of 
$P(\Lambda)$ and a Gaussian distribution gives a quantitative measure 
of how much the motion of the billiard ball in a chaotic 
billiard differs from a random walk. 
\\
(iii) For the integrable rectangular billiard we find a correlation coefficient 
$C \approx 1$, in spectral rigidity $\Delta_{3}(L)$ and in the NNS distribution 
$P(s)$ strong evidence for rigid behaviour and strong correlation 
between neighbour eigenvalues. Such behaviour can be understood from the 
observation that integrability introduces strong correlations in length matrix 
elements. This proliferates to the spectra. 
\\
(iv) In contrast, for integrable quantum systems the NNS distribution generally 
shows Poissonian behaviour with correlation coefficient 
$C=1$ (there are a few exceptions). Thus from the point of view of level spacing 
distributions, a marked difference shows up between the classical 
and the quantum world. In our opinion, such difference is due to the 
group properties$/$symmetries of 
classical trajectories. 
\\
{\it Future directions:}\\
(i) We plan to do numerical studies to investigate if universality also holds 
in chaotic potential systems. 
\\  
(ii) We hope that our findings may contribute to obtain a unified 
description of both, quantum and classical chaos, and help understanding 
why quantum chaos is typically weaker than classical chaos, e.g., 
via an effective quantum action~\cite{Jirari01,Caron01}. 
\\
(iii) The global statistical approach to classical chaos proposed here may help to 
give insight into the problem of ergodicity breaking in Hamiltonian systems 
(e.g., dense packing of discs in the Lorentz gas model~\cite{Gaspard98}).

\section*{Acknowledgments}
JFL is grateful to Prof. LJ Dub\'e for insightful discussions on chaotic dynamics and to O Blondeau-Fournier for discussions and his assistance in performing simulations and analyses presented in section 3.1 and 4.1. HK is grateful to Prof.~Chernov for discussions on central limit theorems in chaotic billiards. HK has been supported by NSERC Canada.


\end{document}